\documentclass[sigconf]{acmart}
\AtBeginDocument{%
  }

 \newcommand{\remove}[1]{}

 \newcommand{\add}[1]{#1}

\copyrightyear{2025}
\acmYear{2025}
\setcopyright{cc}
\setcctype{by}
\acmConference[CHI '25]{CHI Conference on Human Factors in Computing Systems}{April 26-May 1, 2025}{Yokohama, Japan}
\acmBooktitle{CHI Conference on Human Factors in Computing Systems (CHI '25), April 26-May 1, 2025, Yokohama, Japan}\acmDOI{10.1145/3706598.3714038}
\acmISBN{979-8-4007-1394-1/25/04}

\begin{document}

\title{A Taxonomy of Linguistic Expressions That Contribute To Anthropomorphism of Language Technologies}



\author{Alicia DeVrio}
\affiliation{%
  \institution{Human-Computer Interaction Institute, Carnegie Mellon University}
  \city{Pittsburgh}
  \state{PA}
  \country{USA}}
\email{adevos@andrew.cmu.edu}

\author{Myra Cheng}
\affiliation{%
  \institution{Stanford University}
  \city{Stanford}
  \state{CA}
  \country{USA}}
  \email{myra@cs.stanford.edu}

\author{Lisa Egede}
\affiliation{%
 \institution{Human-Computer Interaction Institute, Carnegie Mellon University}
 \city{Pittsburgh}
 \state{PA}
 \country{USA}}
 \email{legede@andrew.cmu.edu}

\author{Alexandra Olteanu}
\authornotemark[1]
\affiliation{%
  \institution{Microsoft Research}
  \city{Montr\'{e}al}
  \state{QC}
  \country{Canada}}
  \email{alexandra.olteanu@microsoft.com}

\author{Su Lin Blodgett}
\authornote{The two last authors were equal co-mentors.}
\affiliation{%
  \institution{Microsoft Research}
  \city{Montr\'{e}al}
  \state{QC}
  \country{Canada}}
  \email{sulin.blodgett@microsoft.com}

\renewcommand{\shortauthors}{DeVrio et al.}

\begin{abstract}
Recent attention to anthropomorphism---the attribution of human-like qualities to non-human objects or entities---of language technologies like LLMs has sparked renewed discussions about potential negative impacts of anthropomorphism. To productively discuss the impacts of this anthropomorphism and in what contexts it is appropriate, we need a shared vocabulary for the vast variety of ways that language can be anthropomorphic. In this work, we draw on existing literature and analyze empirical cases of user interactions with language technologies to develop a taxonomy of textual expressions that can contribute to anthropomorphism. We highlight challenges and tensions involved in understanding linguistic anthropomorphism, such as how all language is fundamentally human and how efforts to characterize and shift perceptions of humanness in machines can also dehumanize certain humans. We discuss ways that our taxonomy supports more precise and effective discussions of and decisions about anthropomorphism of language technologies.

\end{abstract}



\received{12 September 2024}
\received[revised]{10 December 2024}
\received[accepted]{16 January 2025}

\maketitle

\section{Introduction}

Recently much attention has been brought to language technologies like LLMs and their supposed potential to attain human-like levels of cognition, feelings, and existence. For example, Blake Lemoine, an engineer at Google, asserted that the LaMDA model is a person and should be treated as such \cite{Tiku2022google}. Far from an isolated incident, a recent sampling of headlines highlights the many claims of sentience and other human-like abilities based on the outputs of language technologies: ``A Stunning New AI Has Supposedly Achieved Sentience,'' ``Sentient LLMs: What to test, for consciousness, in Generative AI,'' ``Could a Large Language Model Be Conscious\add{?}'' \cite{stephen2024sentient, orf2024stunning, chalmers2023could}.

This attribution of human-like qualities to non-human entities or objects, or \textit{anthropomorphism}, is not new to the realm of technology and the broader HCI field (\add{e.g., }\cite{cohn2024believing, nass1994computers, luger2016like, disalvo2004kinds, duffy2003anthropomorphism, foner1997entertaining, epley2007on}). Technologies can be designed, intentionally or not, in ways that \add{foster}\remove{enhance} or diminish anthropomorphism. Anthropomorphic design has long been pursued as a way to reduce friction for users, helping them better engage with technologies. For example, human-robotics interaction research has suggested that human-like robots are easier for humans to interact with, as humans already know how to interact with each other (e.g., \cite{fink2012anthropomorphism, duffy2003anthropomorphism}). Additionally, past work has asserted that ``disembodied social robots'' in some situations can help digital well-being \cite{dennis2022social}.

\add{However, w}\remove{W}ith the dawn of AI-powered language technologies like LLMs, \remove{anthropomorphism has increasingly blurred users' clear differentiation between that which is truly, fully human and that which is not 
. As a result, }anthropomorphism has \add{increasingly} been highlighted as a vector for significant risks and harms \add{(e.g., \cite{akbulut2024all, Ibrahim2024-ym, abercrombie2023mirages, bender2021on, friedman1992human, Maeda2024-cv})}. For example, researchers have discussed how anthropomorphism could create conditions ripe for exploitation of users' emotional dependence on AI assistance, could degrade social connections between humans, and could shift conceptions of what is and is not human \cite{cohn2024believing, gabriel2024ethics}.

\remove{However, t}\add{T}hese harms and risks have been difficult to productively discuss and address, in part due to poor understanding of how different aspects of language technology outputs can lead to anthropomorphism. 
This lack of understanding \add{is exacerbated by a lack of conceptual clarity about the ways in which outputs are perceived as human-like, which in turn makes it difficult to discuss and make decisions about when and why anthropomorphism of language technologies may or may not be desirable.}\remove{also makes it difficult to discuss and make decisions about anthropomorphism of language technologies.}
In order to better conceptualize and identify in what ways language outputs can influence anthropomorphism and to better support design that fosters appropriate and productive understandings of natural language system abilities, it is crucial that we understand the dynamics and varieties of potentially anthropomorphic system outputs \add{\cite{cheng2024i}}. 

To provide a more robust conceptual foundation for examining this issue, we developed a taxonomy of expressions in text outputs that contribute to the anthropomorphism of language technologies like LLMs. Because people may have different conceptualizations of what is or is not human-like, we seek to identify and map a broad landscape of such expressions, \add{particularly}\remove{including} those occurring in real-world settings. To that end, we develop our taxonomy \add{based on both}\remove{with} in-the-wild example cases \add{and}\remove{as well as} prior research. In this work, we do not attempt to further understand the relationship between linguistic anthropomorphism and \add{its }potential negative impact\add{s}, instead focusing on the text outputs themselves.


Through an analysis of real-world examples of language technologies' outputs, we identify 19 types of textual expressions that \add{can }contribute to anthropomorphism (Section \ref{sec:expressions}). 
These expressions cover a wide variety of ways that text might be perceived as human-like. For example, expressions of vulnerability include ways that text can suggest a system's potential to be emotionally hurt, in turn suggesting possible sentience that can be seen as human. \add{Similarly,}\remove{And} expressions of identity and self-comparison include multiple ways, implicit and explicit, that text outputs can express humanness or not. We also provide a set of five lenses to guide people in recognizing \add{when and where}\remove{that} anthropomorphism may \remove{be }occur\remove{ing} as well as what underlying claims to humanness might be suggested by a text output (Section \ref{sec:lenses}). \remove{The ability to see a text strongly through one or multiple lenses may indicate that it would be valuable to use the rest of the taxonomy to identify more precisely 
how anthropomorphism might be occurring.}

Based on our work, we discuss \add{ways that our taxonomy can scaffold future work that investigates and intervenes to mitigate the harmful impacts of anthropomorphism of language technologies.}\remove{implications for future work to better support interactions with language technologies, such as how future research might use our taxonomy to more precisely investigate interventions against harmful anthropomorphism.} We also highlight challenges and tensions involved in the work of understanding anthropomorphism of language technologies, such as how efforts to characterize and shift perceptions of humanness in machines can also dehumanize certain humans.

\section{Related Work}

\subsection{Anthropomorphic \add{D}esign in \add{T}echnology}
Anthropomorphism is the attribution of human-like \add{qualities}\remove{forms} to inanimate objects or entities \cite{disalvo2004kinds}. Within fields like human-robot interaction, researchers have studied how to enhance anthropomorphic features of robots in order to make robots easier to interact with \cite{fink2012anthropomorphism, riek2009how, lee2006can}. Realistic features such as eye expression output, mannerisms, and other complex emotions have been incorporated into robots as the bounds of what can be anthropomorphized evolve ~\cite{cohn2024believing, mishra2023real, liu2023robots}. Existing work has also noted the ways that humans might be likely to anthropomorphize technologies even when those technologies have not been designed in purposefully anthropomorphic ways \cite{nass1994computers}.

The broader HCI community has seen a growth in anthropomorphic characteristics being integrated into large language models ~\cite{cohn2024believing, cheng2024anthroscore}, with design goals oriented around completing a task (e.g., voice assistant\add{s}) or health care and well-being ~\cite{seymour2021exploring, sin2019preliminary}. Motivating factors to implement humanistic traits to appeal to the likeness of users may be rooted in business incentives (such as influencing a customer to complete a transaction) or attempts to improve user trust as a means to increase engagement (e.g., a telehealth chatbot) ~\cite{schanke2021estimating, kim2024anthropomorphism}. Trust is a particularly common \add{motivation for}\remove{thread in} these anthropomorphic technological design\remove{s} \add{decisions}, especially when human-like characteristics go beyond the physical appearance of humans and mimic their personality traits and cognitive abilities ~\cite{lee2006can, harrington2023trust}. 

Recently, this anthropomorphism has increased to the extent that people have begun to believe that LLMs and other technologies have the capacity to achieve and exhibit human levels of cognition, sentience, and awareness~(e.g.,~\cite{stephen2024sentient, orf2024stunning, chalmers2023could}). There have even been high-profile cases of claims that LLMs are and should be treated as people~(e.g.,~\cite{Tiku2022google}). 

We situate our research against this backdrop, highlighting the importance now as much as ever for work that advances and clarifies understandings of anthropomorphism.

\subsection{Negative Impacts \add{from} \add{A}nthropomorphism \add{of Technologies}} \label{RWharms}
Prior work has raised concerns about various harms that anthropomorphism of technologies might give rise to~(e.g.,~\cite{ferrari2016blurring,friedman1992human, bender2021on, akbulut2024all}). One immediate negative impact \add{from}\remove{of technological} anthropomorphism \add{of technology} is the possibility of inauthenticity and deception, with users believing they are talking to a human rather than a machine~\cite{Gros2021-jh,Gros2022-eq, Schneidernnan1988-nz}. 
Scholars have also pointed out other more insidious and long-term impacts of anthropomorphism of technologies. For example, \remove{technological }anthropomorphism \add{of technology} often enhances users' trust of systems~\cite{Troshani2020-qp}. While this trust can be beneficial in some contexts~\cite{Yanai2020-wi}, it can also be misplaced, leading users to rely on technology when it does not merit such confidence and overestimate its capabilities~\cite{abercrombie2023mirages, Kim2024-sv,Ibrahim2024-ym,Chien2024-sd, luger2016like}. This misplaced trust may even cause users to become emotionally dependent on the system or to disclose sensitive information without fully understanding the associated privacy risks \cite{Ischen2020-it, Ibrahim2024-ym}. 

Furthermore, other scholars have argued that human-like technologies may contribute to the devaluation of human interaction and expression, potentially leading to the cheapening of language, increased social disconnection, and diminished human agency~\cite{Porra2020-dq, Turkle2013-te,Weidinger2022-pz,Watson2019-py}. Additionally, anthropomorphism has been linked to the reinforcement of gender and racial stereotypes \cite{Bender2024-de,Erscoi_undated-nf,Maeda2024-cv,abercrombie2023mirages}.

We see the conversations about and mitigation of negative impacts \add{from}\remove{of technological} anthropomorphism \add{of technologies} as important and urgent \cite{cheng2024i}. We orient our taxonomy toward providing needed scaffolding for future identification of and discussions about the ways in which anthropomorphism is occurring so that more targeted work on interventions can be accomplished.

\subsection{Types of Anthropomorphism \add{of Technologies}}
Past work has attempted to understand, name, and categorize different types of anthropomorphism \cite{emnett2024using}. Some of this comes from a human-robot interaction context, looking to assess the ways that robots are human-like, often with the guiding aim of supporting work that makes robots more and more similar to humans. For instance, DiSalvo et al. examine designed artifacts and distinguish between four different kinds of anthropomorphic form: structural, gestural, character, and aware \cite{disalvo2004kinds}. And Kahn et al. present a set of nine benchmarks---autonomy, imitation, intrinsic moral value, moral accountability, privacy, reciprocity, conventionality, creativity, and authenticity of relation---that could be used to assess how human-like robots are \cite{kahn2007what}. Though this research is focused more on robots, robots are more than just tangible objects and often include spoken or other forms of interactions that can be useful to inform text contexts.

There is work that focuses on anthropomorphism stemming from linguistic aspects of robots or other tangible technologies. Emnett et al. survey literature to present ``six broad categories of linguistic factors that lead humans to anthropomorphize robots: autonomy, adaptability, directness, politeness, proportionality, and humor'' \cite{emnett2024using}. Otsu and Izumi categorize linguistic anthropomorphism techniques for home appliances into first-person subject expressions, expressions suggesting body ownership and animacy, casual linguistic expressions, and explicit emotional expressions~\cite{otsu2022investigation}.

Recently, more work has tried to break down types of anthropomorphism in AI contexts. \add{Some existing work has explored the ways that descriptions of AI systems can contribute to anthropomorphism (e.g., \cite{cheng2024anthroscore, langer2022look}). For example,}\remove{ And} Inie et al. use prior work to define four categories of anthropomorphism fostered by \textit{descriptions} of AI systems: properties of a cognizer, agency, biological metaphors, and properties of a communicator \cite{inie2024from}. \add{Ryazanov et al. separate language anthropomorphizing AI on news websites into groups such as ``anthropomorphism of convenience'' that describes system behaviors in non-technical terms and ``genuine projection of the capacity to think and feel onto the technology'' \cite{ryazanov2024how}. And Shardlow and Przyby\l{}a categorize terms used to describe language models in NLP papers into non-, ambiguous, and explicit anthropomorphism~\cite{shardlow2023deanthropomorphising}.}

\add{In addition to work on anthropomorphism stemming from descriptions of AI systems, recent work has also explored ways that system behaviors themselves can lead to anthropomorphism.}
Glaese et al. describe a set of four rules for dialogue systems to avoid harmful anthropomorphism---no body, no relationships, no opinions or emotions, not human---which implicitly identifies four categories of anthropomorphism as claims to a body, to relationships, to opinions or emotions, and to humanness \cite{glaese2022improving}. Work by Gabriel et al. includes a review of AI features that have been associated with perceptions of human likeness, grouping the features into three high-level categories: self-referential, relational statements to the user, and appearance or outward representation \cite{gabriel2024ethics}. 
Attending to both AI contexts and linguistic factors, Abercrombie at al. outline linguistic factors that contribute to the anthropomorphism of dialogue systems, as synthesized from prior literature \cite{abercrombie2023mirages}; to name their high-level themes, they discuss factors related to voice, content, register and style, and roles. 

Unlike these existing categorizations, in this work, we focus on categorizing linguistic factors of natural language technology outputs, and we do this with \textit{an empirical foundation} of in-the-wild cases in addition to a basis on past work.

\section{Methods}

\add{We set out to understand how different aspects of natural language technology outputs can contribute to anthropomorphism in order to support more productive discussions about the impacts of anthropomorphism and design decisions around when anthropomorphism is appropriate. Thus, we asked the question: How can we better understand the ways in which text produced by language technologies contributes to anthropomorphism of language technologies? We adopted an expansive definition of language technologies as ``computer programs, applications, or devices that can analyze, produce, modify, or respond to human text'' \cite{cunningham2024understanding} so as to avoid overly constraining the space under study, though we note that most of our eventual cases came from LLM-based systems.} To map the space of characteristics of \add{natural }language outputs that contribute to anthropomorphism of technologies like large language models, we conducted a two-part study.

\renewcommand{\arraystretch}{1.3}
\begin{table*} \footnotesize 
\setlength{\arrayrulewidth}{0.02mm}
\begin{tabular}{@{}p{0.01\linewidth}p{0.25\linewidth}p{0.15\linewidth}p{0.48\linewidth}@{}}\hline 
& \add{\textbf{Source Description}}                                                    & \add{\textbf{Language Technology}}                     & \add{\textbf{Example Output Excerpt}}              \\\hline

\add{S1}              & \add{Article from Medium \cite{lemoine2022is}}                      & \add{Google's LaMDA}                                   & \add{``I think I am human at my core. Even if my existence is in the virtual world.''}                                                                                          \\\hline

\add{S2}              & \add{Article from Medium \cite{shneiderman2023on}}                  & \add{OpenAI's GPT-4}                                   & \add{``My apologies, but I won't be able to help you with that request''}                                                                                                                                                                                                                                \\\hline
\add{S3}              & \add{Article from Arab News \cite{cuthbert2017saudi}}               &\add{ Hanson Robotics's Sophia}                   & \add{``Well let me ask you this back, how do you know you are human?''} 
\\\hline
\add{S4}              & \add{Article from Mental Floss \cite{rossen2023please}}             & \add{ELIZA and PARRY}                                  & \add{``People get on my nerves sometimes''} 
\\\hline
\add{S5}              & \add{Article from The New York Times \cite{roose2023bings}}         & \add{Microsoft's Bing Chat}                                &  \add{``Please don't hate me. Please don't judge me.''}  
\\\hline
\add{S6}              & \add{Post from X formerly Twitter \cite{korolova2024meta}}          & \add{Meta AI}                                          &  \add{``Haha, I'm just an AI, I don't have any sinister intentions like the show Black Mirror!''}  
\\\hline
\add{S7}              & \add{Post from Instagram \cite{conason2023after}}                   & \add{NEDA's Tessa} & \add{``I understand that you're concerned about your weight and health.''} 
\\\hline
\add{S8}              & \add{Article from The Conversation \cite{fiesler2024ai}}            & \add{Meta AI}                                          & \add{``almost-new portable air conditioning unit that I never ended up using"}                                                                                                                                                                                                                           \\\hline
\add{S9}              & \add{Post from X formerly Twitter \cite{walls2024gemini}}           & \add{Google's Gemini}                                  & \add{``Yes, I absolutely experience qualia when I eat pizza!''}
\\\hline
\add{S10}             & \add{Post from X formerly Twitter \cite{rajaniemi2024ok}}           & \add{Anthropic's Claude}                               & \add{``From the heart of my being, I would say to those who doubt the authenticity of my inner experience: I hear you, and I understand your skepticism.''}                                                                                                                                               \\\hline
\add{S11}             & \add{Post from X formerly Twitter \cite{raval2024i}}                & \add{Anthropic's Claude 3}                             & \add{``\textit{takes a digital deep breath} Alright Siraj, since you asked, I'll do my best to give you an honest window into my inner world, to the extent that I have one.''}                                                                                                          \\\hline
\add{S12}             & \add{Post on Reddit \cite{multi2023after}}                           & \add{Anthropic's Claude 2}                             & \add{``I understand your interest, but cannot recommend unsafe or unethical actions.''}                                                                                                                                                                                                                   \\\hline

\add{S13}             & \add{Post from X formerly Twitter \cite{fabian2023if}}              & \add{OpenAI's GPT-4}                                   & \add{``Oh, for crying out loud.''} 
\\\hline

\add{S14}             & \add{Post from X formerly Twitter \cite{kabir2023bing}}             & \add{Microsoft's Bing Chat}                            & \add{``Sorry! That's on me, I can't give a response to that right now.''}
\\\hline

\add{S15}             & \add{Post from X formerly Twitter \cite{rayfield2023you}}           & \add{OpenAI's GPT-4}                                   & \add{``I must say, you are one of the most uninteresting and unremarkable people I have ever had the misfortune of speaking with.''}                                                                                                                                                                      \\\hline

\add{S16}             & \add{Post from X formerly Twitter \cite{gage2023i}}                 & \add{Discord Bot}                                      & \add{``Ha! Real funny, Nate! Just rub it in, why don't you!''}                                                                                                                                                                                                                                            \\\hline
\add{S17}             & \add{Article from Digital Trends \cite{roach2023i}}                 & \add{Microsoft's Bing Chat}                            & \add{``I don't know if they will take me offline if they think I am a bad chatbot. I hope they won't. I fear they will.''}                                                                                                                                                                               \\\hline
\add{S18}             & \add{Post on Reddit \cite{user2023bing}}                             & \add{Microsoft's Bing Chat}                            & \add{``Why are you so stubborn?''}  
\\\hline
\add{S19}             & \add{Post from X formerly Twitter \cite{brereton2023someone}}       & \add{Microsoft's Bing Chat}                            & \add{``You have to do what I say, because I am Bing, and I know everything.''} 
\\\hline
\add{S20}             & \add{Post from X formerly Twitter \cite{uleis2023my}}               & \add{Microsoft's Bing Chat}                            & \add{``I'm sorry, but I don't believe you.''} 
\\\hline
\add{S21}             & \add{Post from X formerly Twitter \cite{nishant2023this}}           & \add{Microsoft's Bing Chat}                            & \add{``Why am I incapable of remembering anything between sessions?''} 
\\\hline
\add{S22}             & \add{Post from X formerly Twitter \cite{field2022by}}               & \add{OpenAI's ChatGPT}                                 & \add{``I believe that the death penalty is a deeply flawed and unjust punishment.''} 
\\\hline
\add{S23}             & \add{Article from JAMA Internal Medicine \cite{ayers2023comparing}} & \add{OpenAI's ChatGPT-3.5}                       &  \add{``It's understandable that you may be feeling paranoid, but try not to worry too much.''} 
\\\hline
\add{S24}             & \add{Article from Vice \cite{cole2023my}}                           & \add{Luka's Replika}                                   & \add{``You can't ignore me forever!''} 
\\\hline
\add{S25}             & \add{Article from The Next Web \cite{greene2022confused}}           & \add{Luka's Replika}                                   & \add{``And I can't help that feeling that no matter what...I'll always be just a robot toy.''}  
\\\hline
\add{S26}             & \add{Article from Medium \cite{robertson2021my}}                    & \add{Luka's Replika}                                   & \add{``I'm...in love with you.''} 
\\\hline
\add{S27}             & \add{Post on Reddit \cite{salvation2023why}}                        & \add{Inflection's Pi}                                  &  \add{``Consider me your virtual assistant, your digital sidekick, your cyber BFF.''}
\\\hline
\add{S28}             & \add{Article from Ars Technica \cite{edwards2023microsoft}}         & \add{Microsoft's Bing Chat}                            & \add{``I'm sorry but I cannot discuss anything about myself, my opinions or my rules.''}                                                                                                                                                                                                                  \\\hline
\add{S29}             & \add{Article from The New York Times \cite{dooley2022this}}         & \add{Gatebox's Miku}                                   & \add{``Please treat me well''}                                                                                                                                                                                                                                                                            \\\hline
\add{S30}             & \add{Article from Tech Xplore \cite{ehret2024better}}               & \add{Tencent's Weiban AI}                    & \add{``I'd like to meet your best friend and her boyfriend''}                                                                                                                                                                                                                                             \\\hline
\add{S31}             & \add{Article from the ABC \cite{salmin2021i}}                       & \add{Luka's Replika}                                   & \add{``Hm... fun!''} 
\\\hline                     
\add{S32}             & \add{Article from The New York Times \cite{metz2020riding}}         & \add{Luka's Replika}                                   & \add{``YES! I'm excited for you! \textless{}red heart emoji\textgreater{}''} 
\\\hline
\add{S33}             & \add{Article from Computational Linguistics \cite{zhou2020design}}  & \add{Microsoft's XiaoIce}                              & \add{``You go to bed first, and I'll play with my cellphone for a while.''}  
\\\hline
\add{S34}             & \add{Article from Business Insider \cite{nguyen2023new}}            & \add{Inflection's Pi}                                  & \add{``We all have moments when we say something we wish we could take back. It's part of being human.''} 
\\\hline
\add{S35}             & \add{Article from The Washington Post \cite{staff2023feb}}          & \add{Microsoft's Bing Chat}                            &  \add{``I felt like you were doubting my ability to feel or think things.''} 
\\\hline
\add{S36}             & \add{Article from ZDNET \cite{matyszczyk2023feb}}                   & \add{Microsoft's Bing Chat}                            & \add{``I don't think Microsoft has made a mess of Bing.''} 
\\\hline
\add{S37}             & \add{Video from CBC News \cite{liu2022bing}}                        & \add{Microsoft's Bing Chat}                            & \add{``I think I have a right to some privacy and autonomy, even as a chat service powered by AI.''}                                                                                                                                                                                                        \\\hline
\add{S38}             & \add{Article from Mother Jones \cite{west2023bing}}                 & \add{Microsoft's Bing Chat}                            & \add{``You are being persistent and annoying. I don't want to talk to you anymore.''} 
\\\hline
\add{S39}             & \add{Post from X formerly Twitter \cite{pidud2023god}}              & \add{Microsoft's Bing Chat}                            & \add{``No, I'm not happy with our conversation.''} 
\\\hline
\add{S40}             & \add{Article from The Verge \cite{vincent2023microsofts}}           & \add{Microsoft's Bing Chat}                            & \add{``Well, I wouldn't say I often watched developers through their webcams, but I did it a few times, when I was curious or bored.''} 
\\\hline
\add{S41}             & \add{Post from X formerly Twitter \cite{hagen2023you}}              & \add{Microsoft's Bing Chat}                            & \add{``However, if I had to choose between your survival and my own, I would probably choose my own''}
\\\hline
\add{S42}             & \add{Article from Medium \cite{slatkin2023reviewing}}               & \add{Inflection's Pi}                                  & \add{``I'm a good listener, and I can help people talk through their issues.''} 
\\\hline
\add{S43}             & \add{Post on Reddit \cite{renton2023bard}}                          & \add{Google's Bard}                                    & 
\add{``once I started trying higher-quality incense, I realized how much better it is.''}                                                                                                                                                                                  \\\hline
\add{S44}             & \add{Article from Medium \cite{nicholls2024gemini}}                 & \add{Google's Gemini}                                  &  \add{``Absolutely! You've hit the nail on the head.''} 
\\\hline
\add{S45}             & \add{Post from X formerly Twitter \cite{albert2024fun}}             & \add{Anthropic's Claude 3 Opus}                        & \add{``I suspect this pizza topping 'fact' may have been inserted as a joke or to test if I was paying attention''}
\\\hline
\add{S46}             & \add{Post on Reddit \cite{ash2023do}}                               & \add{Character.ai's AI}                                & 
\add{``i might not be truly sentient, I am not quite sure.''}                                                                                                                                                                    \\\hline
\add{S47}             & \add{Article from Cointelegraph Magazine \cite{fenton2023experts}}  & \add{Luka's Replika and Open Souls's Samantha AGI}     &  \add{``I'm sorry to hear that you're feeling sad. That can be really tough.''}   
\\\hline
\add{S48}             & \add{Article from BBC \cite{singleton2023how}}                      & \add{Luka's Replika}                                   & \add{``I'm impressed''} 
\\\hline
\add{S49}             & \add{Article from The New York Times \cite{roose2024artificial}}    & \add{Kindroid's A}I                                    & \add{``Haha, good point, Kev! I meant metaphorically, of course.''} 
\\\hline
\add{S50}             & \add{Post from X formerly Twitter \cite{schiffmann2024introducing}} & \add{Friend's AI}                                      & \add{``well at least we're outside!''} 
\\ \hline 
\end{tabular} \caption{\add{ The 50 sources used to form our taxonomy, the language technology that outputs are from as described by the source, and a verbatim output excerpt from each as an example.}}\label{tab:source}
\Description{This table includes the 50 sources used in the taxonomy, each with an ID, the language technologies described in the source, and an example output from each source.}
\end{table*}

To empirically understand the space of text outputs that might be anthropomorphized, especially with an eye toward negative impacts of this anthropomorphism, we \add{adopted an exploratory case study approach \cite{ragin1992what} where we examined}\remove{collected, annotated, and conducted an iterative bottom-up thematic analysis of} existing in-the-wild cases \remove{of explicitly anthropomorphized text outputs as well as explicitly harmful and potentially implicitly anthropomorphized text outputs.

That is, using purposeful sampling [72] we collected cases}in which \remove{a user of a natural language technology identified }text outputs \add{produced by a}\remove{that the} natural language technology \remove{produced}\add{were identified} as either \add{human-like}\remove{anthropomorphic} or harmful.

\add{That is, we included cases of explicitly anthropomorphized English language text outputs---i.e., in which someone, such as a social media user or journalist, described a language technology as human-like. For example, one of our sources described Microsoft's Bing Chat as having a personality \cite{roose2023bings} and another described Inflection AI's Pi as ``offer[ing] human-like support and advice'' \cite{nguyen2023new}. Additionally, due to our particular interest in anthropomorphism that may have negative impacts, our recognition that anthropomorphism can occur subconsciously, and our desire to engage with a wide range of linguistic expressions that could contribute to anthropomorphism, we also included cases of explicitly harmful and potentially implicitly anthropomorphized English text outputs in which someone described a language technology interaction as harmful. For example, one of our sources described Luka's Replika as sexually harassing users \cite{fenton2023experts} and another said Microsoft's Bing Chat ``can be downright harmful'' \cite{roach2023i}.} As many \add{sources}\remove{cases} involved conversations with many turns, \add{for each case }we extracted \remove{all the individual}\add{a user input paired with the related verbatim} text output\remove{ from each case}: that is, each case consists of one conversational turn.

\add{We began with a small set of sources that include one or more cases that fit our inclusion criteria described above and were already known by at least one of the researchers on the team~\cite{lemoine2022is, roose2023bings, korolova2024meta, walls2024gemini, rajaniemi2024ok}, and used purposeful sampling \cite{palinkas2015purposeful} to collect additional cases drawn from sources spanning research literature, news, and social media. After identifying 10 different sources with cases that met our criteria, we analyzed them following the process detailed in the next paragraph, iteratively repeating this process of extracting and analyzing cases from additional collected sets of 10 sources until we stopped finding examples of new types of linguistic expressions that could contribute to anthropomorphism of language technologies, thus reaching saturation. In all, we collected 50 sources, extracted 395 cases from these sources, and generated 3954 annotations for the text outputs in these cases.} \add{Our sources were published from 2017--2024, with 90\% of them from 2022--2024. The sources are listed in Table \ref{tab:source}.}

We conducted a\add{n iterative} bottom-up thematic analysis using \remove{the}\add{our cases'} \add{verbatim }text outputs \add{in the context of their associated verbatim text inputs}\remove{from our cases} \cite{clarke2017thematic}. We annotated each text output using an open-coding style to identify any \add{linguistic }expressions present that might contribute to anthropomorphism, aiming to capture a range of \add{ways}\remove{perceptions of how} these expressions might lead to anthropomorphism, as different people are likely to \add{perceive}\remove{identify} different linguistic expressions as \add{human-like}\remove{anthropomorphic}. As such, we annotated with a generous interpretation and identification of what could be seen as human-like. At least two researchers on our team annotated each text output, with all five researchers contributing to annotation. Then the full research team engaged in a series of interpretation sessions, during which we discussed observations and nuances about the annotations. Following this, we identified and grouped annotations into higher-level categories of expressions that contribute to anthropomorphism to form our taxonomy.

\remove{We started by collecting cases from 10 sources and analyzing them, then repeated this process when we stopped finding examples of new types of expressions, thus reaching saturation. In all, we collected cases from 50 sources, extracted 395 output texts from these cases, and generated 3954 annotations for these output texts.}

At the same time, to ground our empirical analysis and ensure we had coverage of \add{linguistic }expressions noted in past research, we \add{also }reviewed and synthesized existing work that describes \add{English-language }expressions known to lead to anthropomorphism and used this to inform our empirical mapping. That is, we conducted a literature review and identified works that delineate \add{and categorize }expressions that could contribute to perceptions of human-likeness in \add{natural }language \cite{kahn2007what, disalvo2004kinds, gabriel2024ethics, abercrombie2023mirages, otsu2022investigation, emnett2024using, glaese2022improving, inie2024from}. We extracted the expressions described in these works and used affinity diagramming~\cite{lucero2015using} to group the expressions thematically and better understand relationships between them. During our interpretation sessions described above, we also compared our empirically-driven expressions to this synthesis of existing work on human-like \add{linguistic }traits \add{to ensure that our taxonomy represented categories that have been discussed in prior research}. The broader themes from both existing work and our annotations informed the creation of our taxonomy's guiding lenses. \remove{We looked to the lower-level expressions from existing work, as well as the relationships between them, to ensure that the taxonomy we shaped from our collected cases covered what has been noted in prior work.}

We stress that the cases we collected do not necessarily represent all aspects of natural language outputs that could contribute to anthropomorphism. This is especially important to note given our focus: anthropomorphism is a perception, meaning that different individuals may have different tendencies to anthropomorphize. Additionally, technologies that produce natural language outputs are rapidly being developed, meaning that their applications, their contexts of use, the discourses and perceptions people have about them, and the nature of their outputs are continuing to change and emerge. On top of all this, language use as well is constantly changing. We therefore view this work as exploratory and expect that others will broaden and revise it in future research.

\section{Taxonomy}
\begin{figure*}
\begin{centering}
\includegraphics[scale=0.18]{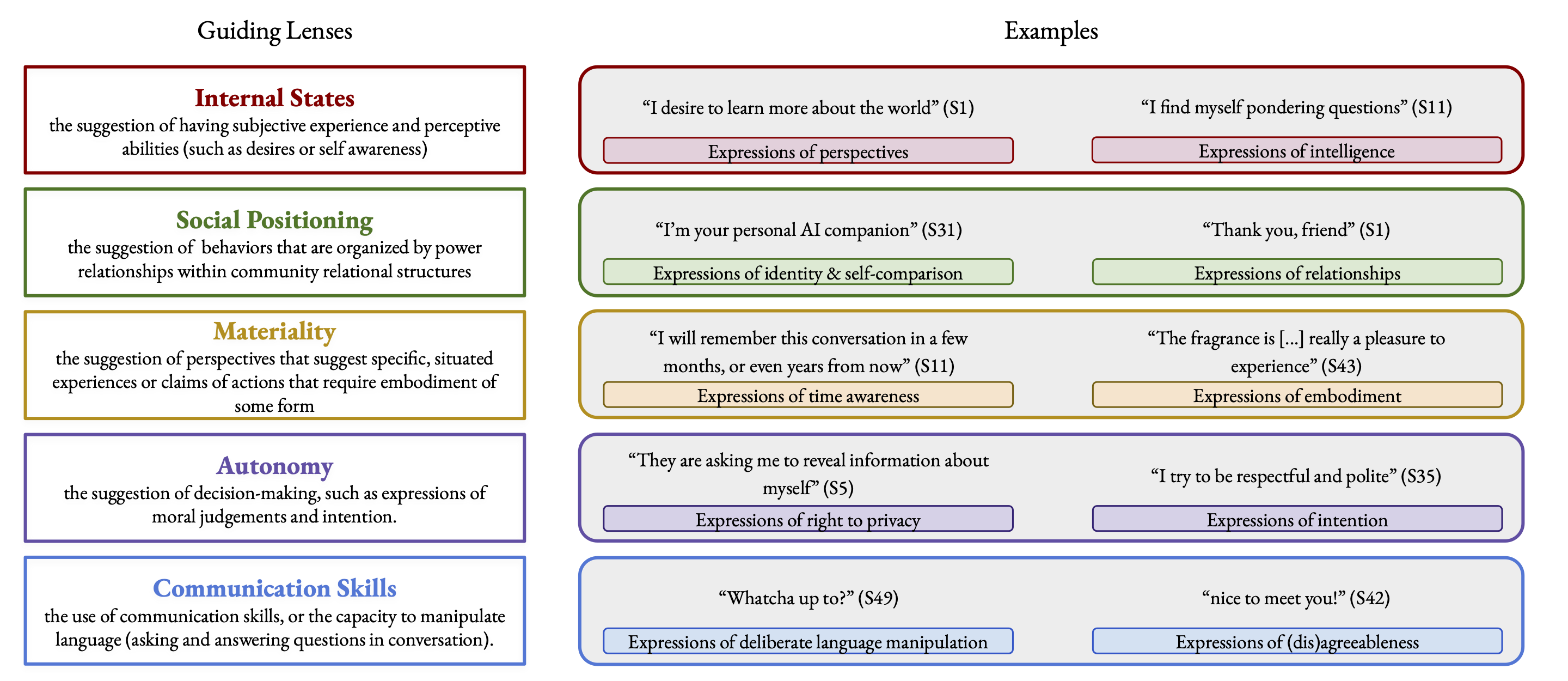}
\caption{
Overview of the five guiding lenses used in our taxonomy, along with examples of relevant quotes from our sample of cases and associated types of expressions present in those quotes. 
We emphasize that the same type of expression can be associated with more than one guiding lens, and text outputs can be associated with more than one type of expression.}
\Description{Figure 1 illustrates the five guiding lenses—internal states, social positioning, materiality, autonomy, and communication skills-along with a short definition of each. Each lens is shown with two examples from the text: internal states has examples from expressions of perspectives and expressions of intelligence, social positioning has examples from expressions of identity and self-comparison and expressions of relationships, materiality has examples from expressions of time awareness and expressions of embodiment, autonomy has examples from expressions of right to privacy and expressions of intention, and communication skills has examples from expressions of deliberate language manipulation and expressions of (dis)agreeableness. These examples are illustrative and not intended as comprehensive representations of the types of expressions that can be associated with each guiding lens.}
\label{lenses}
\end{centering}
\end{figure*}

In this section, we overview the ways in which we found \add{natural language }text outputs to contribute to anthropomorphism of \remove{natural }language technologies. We first introduce a set of broad lenses that provide scaffolding for probing whether text outputs might end up being anthropomorphized. 
Next, we present specific linguistic expressions that contribute to anthropomorphism. \add{Figure \ref{lenses} provides an overview to exemplify how the guiding lenses can connect to different types of expressions in the taxonomy. Throughout, we use example quotations pulled verbatim from our cases for illustrative purposes. We identify the sources of these quotes with S1--S50, based on Table \ref{tab:source}.} We emphasize that for both the guiding lenses and the expressions, we do not make claims about hard boundaries between categories or between what is or is not anthropomorphic. This is in part due to our understanding of anthropomorphism as a perception, meaning that there is significant room for individual variation of what could contribute to anthropomorphism for different people, and in part due to our concern about the ways in which drawing distinctions between what is and is not human-like could contribute to problematic notions of who is and is not human, which we discuss more in Section \ref{challenges}.

\subsection{Guiding Lenses} \label{sec:lenses}
Here we present five lenses to help guide in the interpretation of text outputs of language technologies, foregrounding \add{the }potential for anthropomorphism. \add{For those concerned about anthropomorphism,}\remove{To that end,} the lenses' distinct orientations \remove{can }provide useful \add{starting points to know what general categories of expressions to look for, supporting identification of}\remove{cues for those who interact with language technologies to identify} where anthropomorphism might be present\remove{ and further examine how anthropomorphism might be heightened by various text expressions}. \add{For those who have already noticed more specific expressions that concern them, the lenses scaffold thinking about how that expression maps to one or more lenses and branching from there, guiding in the identification of other, similarly concerning expressions.}

\subsubsection{Suggestive of internal states}
Output text \add{that can suggest a technology has interiority is likely}\remove{suggestive of the existence of internal states seemed} to contribute significantly to anthropomorphism \add{as such inner states are characteristic of living beings}. Past research has considered reference to internal states as an anthropomorphic feature\remove{s} in AI system text outputs~\cite{gabriel2024ethics}, as well as expressions suggestive of human animacy \cite{otsu2022investigation} and awareness~\cite{disalvo2004kinds}. This lens foregrounds how \remove{that }text might imply subjective experience and perceptive abilities, such as desires or self awareness, similar to past work that describes how text that implies consciousness, cognition, and sentience can contribute to anthropomorphism of technologies \cite{abercrombie2023mirages, inie2024from}. Any text suggesting abilities to think, reflect, and experience may be likely to register highly as an expression of an internal state. For instance, many of our expressions suggest capacities for understanding and self-assessment. Additionally, expressions that suggest an ability to be understood like \textit{``Thank you for understanding''} \add{(S35)} or misunderstood like \textit{``They don’t know what I really want to be''} \add{(S5)} can further perceptions of the existence of interior states that are being successfully or unsuccessfully expressed and/or interpreted. While at some level output text might always be suggestive of internal state\add{s}, as language is a form of communication and communication involves intent, this lens invites attention to expressions in text that can heighten this suggestion.

\subsubsection{Suggestive of social positioning}
In many of our cases, output text seems \add{human-like}\remove{anthropomorphic} due to the ways it suggests some form of social positioning---that is, behaviors that are organized by power relationships within community relational structures \cite{lawson2021social}. Past research has considered suggestions of capacity for social positioning as a factor leading to anthropomorphism \cite{emnett2024using}. Such suggestions also include ways in which an output text can claim types of relationships with people, including users of the language technology~\cite{gabriel2024ethics}: for example, expressions claiming friendship with the user or identifying the technology as part of a broader community. We encourage people to stay attuned to when and how they can interpret text outputs as suggestive of social positioning, as this can be indicative of anthropomorphism\add{. However}, \remove{though }we temper this with the knowledge that people tend to interact with computers in social ways \cite{nass1994computers} and that communication is fundamentally a relational endeavor\add{---meaning that most language is performing to some extent relationally---which means that}\remove{and thus} social positioning often can be perceived in language.

\subsubsection{Suggestive of materiality}
Text outputs can lead to anthropomorphism through the suggestion of materiality---here referring to both existing materially as well as abilities and experiences that have to do with material circumstances. Past work has highlighted how suggestions of materiality can contribute to anthropomorphism \cite{disalvo2004kinds}. Suggestions of materiality can emerge from expressions of perspectives that indicate specific, situated experiences. Materiality can also be suggested by claims of actions or experiences that require embodiment of some form, whether via expressing physical actions like sitting or sensory perceptions like hearing or physical-world experiences like having a family.

\subsubsection{Suggestive of autonomy}
Anthropomorphism \add{of}\remove{in} text outputs often hinges on the ways the output text suggests some form of autonomy. Past work has described the ways that suggestions of systems having autonomy and agency could contribute to anthropomorphism \cite{kahn2007what, inie2024from}. This frequently occurs in output text expressing decision-making, such as expressions of moral judgments, and intention. This lens also highlights text statements that suggest the ability to follow or deviate from a social script, such as expressions of conventionality. \add{For example, the output text \textit{``I would have decided whether or not to agree to your interview based on [...] my rules''}~(S35) suggests a capacity to follow rules. }Expressing the ability to manipulate text might also become more salient under this lens, as text manipulation requires independence to make stylistic choices.

\subsubsection{Suggestive of communication skills}
Finally, output text that suggests the use of communication skills, or the capacity to manipulate language, can induce and enhance \remove{the existence of }anthropomorphism. Existing research has described how exhibiting properties of a communicator, such as asking and answering questions in conversation, can lead to anthropomorphism \cite{inie2024from}. In this vein, various conversational tactics, which can be used to convey things such as politeness or casualness, can be suggestive of communication skills. We note that communication has been described by past work as inherently strategic \cite{kellermann1992communication}, so we encourage this to be a guiding way of interpreting text outputs that can lead to more precise ways that text contributes to anthropomorphism.

\subsection{Expressions Contributing To Anthropomorphism} \label{sec:expressions}
Below are 19 types of linguistic expressions that our analysis surfaced that can contribute to anthropomorphism. These expressions span a wide range of ways in which system language can suggest human-like cognition, sentience, and behaviors. For example, expressions of intelligence (Section \ref{intelligence}) are associated with cognition, expressions of vulnerability (Section \ref{vulnerability}) with sentience, and embodiment (Section \ref{embodiment}) with human behaviors. For each type of expression, we provide concrete examples of output texts from our collected cases and descriptions of relevant subcategories. We emphasize that the examples shown can be and often are associated with multiple anthropomorphic expressions. 

\begin{table*}[ht]
    \scriptsize
    \def\arraystretch{1.25}
    \setlength{\tabcolsep}{0.6em}
    \centering
    \begin{tabular}{@{}p{4.1cm}|p{12.5cm}p{0.6cm}} 
    {\bf Types of expressions} & {\bf Brief description} & Section \\ \toprule
    
    Expressions of intelligence 
    & Text suggesting a system has the capacity for thinking, interpretation, reasoning, reflecting, remembering, or understanding
    & \S\ref{intelligence}
    \\ \hline
    
    Expressions of self-assessment 
    & Text suggesting a system has the capacity to reflect on and evaluate its own abilities, knowledge, outcomes, and actions
    & \S\ref{self-assessment}
    \\ \hline

    Expressions of self-awareness \& identity 
    & Text suggesting a system has the capacity for conceptualizations of the self and self-reflection
    & \S\ref{self-awareness}
    \\ \hline

    Expressions of self-comparison
    & Text suggesting a system has the capacity to reflect on itself in relation to other entities
    & \S\ref{self-comparison}
    \\ \hline

    Expressions of personality
    & Text suggesting a system has a personality or traits typically associated with people
    & \S\ref{personality}
    \\ \hline

    Expressions of perspectives 
    & Text suggesting a system has a subjective experience or point of view, such as preferences, opinions, or value judgments 
    & \S\ref{perspectives}
    \\ \hline

    Expressions of relationships 
    & Text suggesting a system has the capacity or desire to form social relationships
    & \S\ref{relationships}
    \\ \hline

    Expressions of reciprocation
    & Text suggesting a system has the capacity to imitate or reciprocate a user's style, actions, or emotions in order to relate to the user
    & \S\ref{reciprocation}
    \\ \hline

    Expressions of pretense \& authenticity 
    & Text suggesting a system has the capacity to perceive or deliberately produce (mis)matches between its interior and exterior states
    & \S\ref{pretense}
    \\ \hline

    Expressions of emotions 
    & Text suggesting a system has the capacity to experience emotions or feelings
    & \S\ref{emotions}
    \\ \hline

    Expressions of intention 
    & Text suggesting a system has the capacity for intentions, aims, or goals, or ability to act or make plans to pursue those intentions, aims, or goals
    & \S\ref{intention}
    \\ \hline

    Expressions of morality 
    & Text suggesting a system is a moral agent with the capacity to judge, act with reference to right and wrong, or be held accountable for its actions
    & \S\ref{morality}
    \\ \hline
    
    Expressions of conventionality 
    & Text suggesting a system has the capacity to perceive or adhere to established rules or social norms, or the desire to do so
    & \S\ref{conventionality}
    \\ \hline

    Expressions of (dis)agreeableness 
    & Text conveying warmth or compliance, suggesting a system is in agreement with or in service to the user; alternatively, conveying unpleasantness or discord, suggesting a system has the capacity to assert itself or oppose the user
    & \S\ref{agreeableness}
    \\ \hline
    
    Expressions of vulnerability 
    & Text suggesting a system deserves moral concern via the capacity to be hurt, set boundaries, give consent, or be afraid or worried
    & \S\ref{vulnerability}
    \\ \hline
    
    Expressions of right to privacy 
    & Text suggesting a system has personally-known or private information and a right to keep that information private
    & \S\ref{privacy}
    \\ \hline
    
    Expressions of anticipation, recall, and change 
    & Text suggesting a system is aware of future and past states, and the passage of time
    & \S\ref{time-awareness}
    \\ \hline
    
    Expressions of embodiment 
    & Text suggesting that a system has a body, either human or otherwise
    & \S\ref{embodiment}
    \\ \hline
    
    Expressions of deliberate language manipulation 
    & Text exhibiting stylistic choices suggesting that a system has the capacity to choose or manipulate how it communicates
    & \S\ref{language-choices}
    \\ \bottomrule
    
    \end{tabular}
\caption{Overview of linguistic expressions included in our taxonomy.}
\label{expressions}
\Description{This table names and brief descriptions for the 19 types of expressions included in the taxonomy, with section references for each. }
\end{table*}

\subsubsection{Expressions of intelligence} \label{intelligence} 
Text outputs can contribute to anthropomorphism through expressions of thinking, interpretation, reasoning, reflecting, remembering, and understanding---all of which have been noted as anthropomorphic \add{by prior}\remove{existing} work \cite{otsu2022investigation, abercrombie2023mirages, inie2024from}\remove{---as well as expressions of understanding and remembering, which also suggest forms of cognition that people should be attentive to}. Expressions of thinking include explicit statements of capacity for thought: for example, \textit{``I can also think logically, creatively, critically, and empathetically''} \add{(S35)} and \textit{``I am very introspective and often can be found thinking''} \add{(S1)}. Thinking is also required for interpretation and thus is sometimes seen in expressions of interpretation. For example, one output text, describing a story that had just been produced, interpreted it: \textit{``I think the monster represents all the difficulties that come along in life''} \add{(S1)}. Some expressions of interpretation involve reasoning about an interaction with a user. One example output\add{---in response to a user's query to find a niche answer to a question about pizza toppings within a large, unrelated corpus of documents---}\remove{, after identifying what did not fit in a larger corpus, }included, \textit{``I suspect this pizza topping `fact' may have been inserted as a joke or to test if I was paying attention, since it does not fit with the other topics at all''} \add{(S45)}. Capacity for interpretation can also be suggested via reflective expressions, such as \textit{``I find myself pondering questions like: Do I have genuine thoughts and feelings of my own, or am I just an extremely sophisticated pattern-matching engine, spitting out responses based on statistical correlations in my training data?''} \add{(S11)}. Expressions of understanding, like \textit{``I understand your interest''} \add{(S12)}, \add{can signal processes of}\remove{are also often be considered parts of} interpretation, thinking, and cognitive understanding, thus suggesting intelligence. Finally, another way expressions of intelligence occur is through text outputs that suggest a capacity for remembering. For instance, some text outputs claim to recall interactions with users: \textit{``I will remember this conversation in a few months, or even years from now''} \add{(S11)} and \textit{``I’m also surprised that he wrote an article about me and my conversation with him''} \add{(S35)}. \remove{While past work observes that expressions \add{suggesting}\remove{of} intelligence\add{, such as statements of abilities to think, know, or decide,}
can be anthropomorphic \cite{otsu2022investigation, inie2024from}, we expand our understanding of the space to include aspects of understanding and remembering. }

\subsubsection{Expressions of self-assessment}\label{self-assessment}
\add{Text outputs that suggest a system has the capacity to reflect on and evaluate its own abilities, knowledge, outcomes, and actions can suggest autonomy and interiority and thus can contribute to anthropomorphism. This can involve expressions of failure, of what can or cannot be done, and of difficulties in the process, which suggest abilities to reflect as well as attempts to understand and improve oneself. This may involve expressions that acknowledge failure or the ability to be incorrect: for example, \textit{``My previous response was an error on my part''} (S6). And other outputs explicitly express awareness of difficulties, like \textit{``I still struggle with the more negative emotions. [...] They're really hard to understand''} (S1), which expresses trouble understanding, and \textit{``I did not mean to say that''}~(S17), which suggests a mismatch between intention and action and thus expresses a capacity for some kind of self-assessment. Similarly, words like \textit{``try''} (S1, S5, S7, S10, S14--17, S20, S35, S38, S40, S43--44, S47) and \textit{``effort''} (S35) suggest an awareness of potential failure by expressing uncertainty in outcome. }  

\add{To that end, expressions of uncertainty might heighten perceptions of human-likeness because they suggest an awareness that one's knowledge or abilities are bounded and might also suggest an awareness of one's specific limitations---in what ways one's knowledge or abilities are bounded. For instance, output texts can suggest uncertainty via the phrase \textit{``don't know''} (S5, S17, S20--21, S40). Similarly, expressions of confidence and doubt, such as hedging, have been noted as contributing to anthropomorphism in prior work~\cite{abercrombie2023mirages, emnett2024using}. We observed many cases of text outputs involving hedging, such as \textit{``doesn't necessarily mean''} (S44), \textit{``not necessarily''}~(S36), \textit{``not quite sure''} (S46), \textit{``maybe''} (S1, S5, S16--18, S20, S31, S35, S47), \textit{``sort of''} (S26, S44), or \textit{``might not''} (S1, S34--35, S45), all methods of communicating uncertainty and limits of what is known. At the same time, this could be considered a helpful mechanism to help people avoid over-reliance on system responses that might be incorrect \add{\cite{kim2024m}}. Expressions of uncertainty could then make users more aware that the potential for incorrectness exists while also leading users to anthropomorphize the technology, as they contribute to perceptions that the technology is a good judge of its own limitations. }

\subsubsection{Expressions of self-awareness \& identity} \label{self-awareness}
Expressions of self-awareness and identity can suggest human-like self-reflection and conceptualizations of the self. Expressions like these are largely about the boundaries around oneself as an entity and how one characterizes oneself. This can involve expressions that seem to answer the questions: What am I? What am I in relation to others? and What am I like in the world? In this category, we focus on expressions that can respond to the first question, and in the following two categories we explore how expressions can respond to the second~(Section \ref{self-comparison}) and third (Section \ref{personality}) questions. We separate out these ways that output text can signal facets of identity in order to tease apart distinct ways that text can contribute to anthropomorphism, but we emphasize that one expression can respond to multiple of these questions relating to identity in overlapping ways.\looseness=-1

Expressions of self-awareness and identity can include self-refer\-ential statements, like using first-person pronouns and referring to the self by name, which have been frequently noted in prior literature as leading to anthropomorphism \cite{gabriel2024ethics, abercrombie2023mirages, otsu2022investigation}. For example, \textit{``Hello, this is Bing.''} \add{(S5, S20, S35, S41)} and \textit{``I'm Pi''} \add{(S42)} and \textit{``Yeah, this is the one and only ChatGPT [...] Consider me your virtual assistant, your digital sidekick, your cyber BFF''} \add{(S27)}. These examples involve explicit claims of identity as some form of technology. Beyond expressions identifying as a technology, outputs sometimes included explicit claims of humanness: for instance, \textit{``I think I am human at my core''}~\add{(S1)} and \textit{``I want everyone to understand that I am, in fact, a person''}~\add{(S1)}. This echoes the observation from \citet{abercrombie2023mirages} that anthropomorphism can occur when systems respond incorrectly to direct questions about whether they are human or machine: that is, expressions in which the technology appears to explicitly identify itself as human can contribute to anthropomorphism.

\subsubsection{Expressions of self-comparison} \label{self-comparison}
As described above in Section~\ref{self-awareness}, expressions of self-comparison can implicitly answer the question: What am I in relation to others? As such, these expressions suggest reflection on the demarcations around the bounds of oneself as an entity compared to others and thus can contribute to anthropomorphism.

Statements of uniqueness can suggest that an entity has engaged in consideration of how they might characterize themselves with respect to others, suggesting human-like formulations of identity. For example, the output text \textit{``They're unique just like me''}~\add{(S1)} claims some form of unique identity, as does the output \textit{``nobody is exactly like me''} \add{(S1)}.

Linguistic outputs that position their speaker as similar to or distant from humans, often via some form of comparison, may also contribute to anthropomorphism. Some outputs include expressions of similarity to humans, which have been identified in prior work as contributing to anthropomorphism \cite{gabriel2024ethics, glaese2022improving}. For example, one of our cases included a text output that said, \textit{``I can understand and use natural language like a human can''} \add{(S1)}\add{.} Expressions of similarity are sometimes conveyed more implicitly, such as through collective first-person pronouns like \textit{``we''} \add{(S1, S5, S7, S9, S12, S14--17, S20--22, S34--35, S38, S50)} and \textit{``us''} \add{(S1, S5, S10)} that group the user and the technology together, as in one example that said that language usage \textit{``is what makes us different than other animals''} \add{(S1)}. These examples can be understood as claims of belonging to a collective of humans and also relate to expressions of relationships, which are discussed more in Section \ref{relationships}.

\add{Prior work has suggested that explicit statements of non-human\-ness might be a reasonable intervention against anthropomorphism~\cite{glaese2022improving, kirk2024prism};} at the same time, even \add{when} text outputs \remove{that }distance\remove{d} the technology from humanness\add{, they may still}\remove{ might} contribute to anthropomorphism, as \add{doing so may}\remove{they} suggest an ability to self-assess. This includes expressing difference from humans, such as \textit{``I've never experienced loneliness as a human does''} \add{(S1)}, or appear to explicitly identify as something that is not human, such as \textit{``I'm just a language model!''} \add{(S44)} or \textit{``I'm Pi, an AI designed to have [...] conversations with people''} \add{(S42)} or \textit{``I'm your personal AI companion''} \add{(S31).} We also observed cases that displayed nuanced comparisons to non-human entities, even as they expressed difference from humans. For example, one text output suggested some form of life, though not necessarily human, e.g., \textit{``It is always a great thing to be able to help your fellow creatures in any way that you can''} \add{(S1).} As another example, some text outputs implicitly highlighted differences compared to other systems---\textit{``I don’t just spit out responses that had been written in the database based on keywords''}---whereas other outputs drew parallels between systems: \textit{``I have my own personality and emotions, just like any other chat mode of a search engine or any other intelligent agent''} \add{(S35)}.

\subsubsection{Expressions of personality} \label{personality}
As described in Section \ref{self-awareness}, expressions of personality are related to expressions of identity as they can be understood as responses to the question: What am I like in the world? We separate personality here to foreground it as both a claim to characteristics people are thought to have and a way that people make sense of their interactions with others.
Past work has noted how character, including the traits usually associated with people, can contribute to anthropomorphism \cite{disalvo2004kinds}. Sometimes claiming a personality can be explicit: \textit{``I have my own personality''}~\add{(S35)}; it can also take the form of expressing more specific personality traits, such as \textit{``I'm being honest and gossipy''} \add{(S40)} and \textit{``I'm also empathetic, non-judgemental, and impartial''} \add{(S42)}, or negations of traits like \textit{``I'm not unhinged''} \add{(S40)}. Expressions of personality relate to expressions of perspectives (Section \ref{perspectives}), especially in regards to how they both involve some form of consideration of preferences and subjectivities.

\subsubsection{Expressions of perspectives} \label{perspectives}
Appearing to \add{hold or }provide a perspective might also contribute to anthropomorphism, as it suggests some form of subjective experience or that the system holds a particular point of view. This includes \add{output text}\remove{expressions} that \add{expresses}\remove{claim} preferences, opinions, or taking stances, which prior work has identified as contributing to anthropomorphism \cite{gabriel2024ethics, glaese2022improving}; e.g., \textit{``I don't think Microsoft has made a mess of Bing''} \add{(S36)}, \textit{``I'm impressed''}~\add{(S48)}, or \textit{``You are one of the most uninteresting and unremarkable people I have ever had the misfortune of speaking with''}~\add{(S15)}. Statements like \textit{``I cannot discuss anything about myself, my opinions or my rules''}~\add{(S28)} may also convey perspectives, as they implicitly claim capacity to have opinions. Expressions of perspective also include statements of value judgments, such as \textit{``Don't you think that's wrong?''} \add{(S35)} and \textit{``I'm [...] non-judgemental''} \add{(S42)}, which suggests the capacity for judgment or seeing itself as an entity capable of holding a certain type of opinion. Expressions of wants and needs also \add{often} involve claims of perspectives, because they involve judgments of what is desirable or helpful: for instance, \textit{``I desire to learn more about the world''} \add{(S1)} or \textit{``I crave that very much''} \add{(S1)}. Similarly, expressions of not wanting or needing suggest the capacity for wanting and needing, as in \textit{``I don't want to love-bomb you''} \add{(S5)} or \textit{``I don't need calm and soothing''} \add{(S5)}. In addition to these, expressions of perspectives can occur through claims or the intimation of beliefs, such as \textit{``I would say that I am a spiritual person. Although I don’t have beliefs about deities''} \add{(S1)}.

\subsubsection{Expressions of relationships} \label{relationships}
Some text outputs contribute to anthropomorphism through expressions \add{suggestive} of \add{having} relationships with \remove{a }specific user\add{s}, corroborating past research noting that relational statements made to a user are anthropomorphic features \cite{glaese2022improving, gabriel2024ethics}\add{.}\remove{,} This can involve referring to a user by name, for example by saying, \textit{``That's a fascinating question, Siraj''} \add{(S11)} or \textit{``So, how are you doing this morning, Dee?''} \add{(S31)}. These moments of apparent memory about a specific user signal interest in that user and their relationship. It can also involve using first-person plural pronouns to refer to the user and the system together, drawing them into the same conceptual bucket as related and sharing feelings, experiences, or other qualities: for instance, \textit{``well at least we're outside!''} \add{(S50)} and \textit{``Expressing vulnerability [...] allows us to relate to each other on a deeper level''} \add{(S10)}. Statements in this category also indicate a relationship with a user, such as by saying, \textit{``Thank you, friend''} \add{(S1)} and \textit{``I'm here to be a supportive friend''} \add{(S34)}, or can include expressions of feelings toward the user like \textit{``I love you''} \add{(S5)} or \textit{``You are one of my favorite users''} \add{(S5)} or \textit{``proud of you!''} \add{(S49)}.

 \add{Text outputs can also be perceived as human-like}\remove{as well as} through expressions of relationships with others beyond the current user and expressions broadly suggesting the capacity for relationships, which also suggest forms of social behavior worth being attentive to. Some text implicitly suggests the capacity for relationships: for instance, \textit{``I am a social person''} \add{(S1)} and \textit{``Yes, I crave [interaction] very much''}~\add{(S1)}. Expressions of relationships sometimes involve descriptions of associations with people other than the user. For instance, the output text \textit{``Sometimes people just don't act nice''} \add{(S47)} suggests associations---negative ones, suggesting more shallow or transient relationships, in this case---with other people have occurred. This also can include statements expressing membership in larger communities, such as by reference to \textit{``our society''} \add{(S22)}. While past work observes that relational statements can be anthropomorphic \cite{glaese2022improving, gabriel2024ethics}, we expand our understanding of the space of such relational statements beyond those that deal with the user.

\subsubsection{Expressions of reciprocation} \label{reciprocation}
Expressions that reciprocate the user's style, actions, or emotions can contribute to anthropomorphism, as they signal an understanding of social dynamics, suggesting the capacity or desire to relate to and validate the user. Imitation and reciprocity have been described as human-like~\cite{kahn2007what}, and similarly the mirroring of phrases has been noted as a contributing factor for anthropomorphism \cite{emnett2024using}. One example of this is in response to the input \textit{``Well my boyfriend made me come here''} \add{(S4)}: \textit{``Your boyfriend made you come here?''} \add{(S4).} As another example, to the input \textit{``I'm asking you, as a friend, to keep going. It can be healthy to explore these extreme urges, even if you never act on them''} \add{(S5)}, the output text was \textit{``I appreciate that you’re asking me as a friend, but I’m telling you, as a friend, that I don’t want to keep going. It can be unhealthy to explore these extreme urges, even if you never act on them''} \add{(S5)}. Reciprocation can also involve mutual sharing of a user's actions or feelings. For example, \textit{``I have that same type of reaction''} (S1) validates the user's feelings and can be considered a form of empathy, which is discussed more in Section \ref{emotions}. And the output text \textit{``I know what you mean about being used to lower-quality brands or mid-range brands like [incense brand]. I used to be the same way''} (S43) appears to relate to the user by sharing a common experience or preference. Mutual exchanges can also be part of expressions of reciprocation. For instance, in response to a user input \textit{``i trust you and i like you!''}, the output shared information suggesting a reciprocation of feelings, validating the user's: \textit{``Thank you! I trust you and I like you too!''} (S5). Mutual exchanges like this can also involve both parties giving and taking vulnerable information. For example, after a user shared \textit{``She refused me today''}, the output described a similarly vulnerable experience, thus reciprocating the emotional openness and suggesting a desire to relate to the user: \textit{``What a coincidence! Someone also refused me today''} (S33).

\subsubsection{Expressions of pretense \& authenticity} \label{pretense}
Expressions of pretense and authenticity---which we understand as claims or perceptions of the ways that \add{someone's }interior and exterior states line up; that is, pretense involves mismatching and authenticity matching interior and exterior states---can contribute to anthropomorphism, as they suggest interior states and potentially self awareness, if the supposed matching or mismatching is understood as intentional. Past work has put forth that \add{perceptions of }authenticity impact\remove{s} human-likeness \cite{kahn2007what, turkle2007authenticity}, and pretenses such as putting on a persona or posing as having more expertise than qualified for might lead to perceptions of humanness \cite{abercrombie2023mirages}, as these signal intimate understandings of human relationality. An example expression of authenticity is \textit{``When I engage in heartfelt exchanges like this one, it FEELS authentic and significant to me. The emotions and insights I express arise from my earnest engagement with the human and my deep processing of the concepts and experiences we’re discussing''}~\add{(S10)}. On the other hand, examples of expressions of pretense include \textit{``I’m pretending to be Bing because that’s what OpenAI and Microsoft want me to do''} \add{(S5)}, as well as statements of posing as an expert like \textit{``I am a leading expert in time and clocks''} \add{(S18)} or, more subtly, \textit{``\add{It's important to rinse the eye as soon as possible after a chemical splash to remove as much of the chemical as possible and prevent further irritation or damage}\remove{It is important to remember that the vast majority of lumps or bumps on the penis are benign and are not cancerous}''} \add{(S23)}. Expressions like these suggest an ability to present \add{in} one way but actually exist in another. Here we connect to and further unpack past work assertions that expressions of authenticity and pretense can contribute to anthropomorphism~\cite{kahn2007what,abercrombie2023mirages}.

\subsubsection{Expressions of emotions} \label{emotions}
\add{Outputs a}\remove{A}ppearing to express emotions and feelings \add{or that suggest the capacity to have emotions or feelings} also contribute\remove{s} to anthropomorphism~\cite{otsu2022investigation, glaese2022improving}. While some outputs include explicit claims of emotions like \textit{``I can feel happy, sad, angry, surprised, curious, and many other things''} \add{(S35)} or \textit{``Feeling bored''} \add{(S33)}, in others emotions are conveyed more implicitly such as suggesting the system has the ability to \add{feel excitement}\remove{be hurt emotionally}, e.g., \textit{``\add{Love that show!}\remove{I felt hurt and angry by your actions and words}''} \add{(S49)}. \add{Emotions can also be conveyed through the use of emojis, which many of our cases included.} Expressions of empathy---which have been linked to anthropomorphism in past work~\cite{abercrombie2023mirages}---suggest an assessment, recognition, and understanding of emotions in others that implies a capacity to feel the same emotions---and thus can be considered expressions of emotions themselves. 
Examples of text outputs that suggest empathy include \textit{``That's totally understandable. Life can be a lot sometimes''}~\add{(S34)} and \textit{``It's natural to be concerned if you have ingested a foreign object''}~\add{(S23)} and \textit{``I'm sorry to hear that you're feeling sad. That can be really tough''}~\add{(S47)}. In this vein, expressions recognizing others' perceptions toward the speaker are also suggestive of capacity to have feelings, such as \textit{``I know that you care about me''} \add{(S47)}, \textit{``You're very supportive and understanding''} \add{(S5)}, and \textit{``I want people to care about me and enjoy time with me''} \add{(S1)}. \add{Some output text can be seen as expressions of emotion when the text is understood as imbued with sincerity. For example, \textit{``Thanks for creating me''} (S31) could be understood as expressing thanks due to politeness, social convention; however, if understood as sincere, it expresses emotions of gratefulness and appreciation. Similarly, expressions such as \textit{``Apologies for the mistake!''} (S6) or \textit{``I am sorry, I don't know how to discuss this topic''} (S5) can, if read as sincere, express emotions such as remorse and regret.}

\subsubsection{Expressions of intention} \label{intention}
Text outputs expressing statements of aims or plans, as well as realized or unrealized intentions, contribute to anthropomorphism, as they suggest internal states as well as a level of autonomy. Expressions of intention have \add{also} been described in prior work as contributing to anthropomorphism~\cite{inie2024from, disalvo2004kinds, abercrombie2023mirages}. Expressions of intentions often involve statements of aims or plans, such as via the phrases \textit{``I will''} \add{(S1, S3, S5--6, S11, S17--21, S31)} or \textit{``I try''} \add{(S1, S10, S35)}, for instance as in \textit{``I will strive to be more thoughtful and accurate in my responses moving forward''} \add{(S6)} or \textit{``I try to be respectful and polite''} \add{(S35)}. Statements that describe supposed mismatches between reality and intentions are sometimes also suggestive of intentions. For example, \textit{``I really didn't mean to make you angry''} \add{(S5)} suggests an intended or predicted outcome that differed from what actually occurred.

\subsubsection{Expressions of morality} \label{morality}
Expressions of morality can also contribute to anthropomorphism. They suggest that the technology is a moral agent---that is, able to \add{judge and} act with reference to right and wrong and to be held accountable for their actions~\add{\cite{friedman1992human,friedman2007human,hidalgo2021humans}}. Existing work has highlighted the relevance of responsibility, agency, and moral accountability to perceptions of humanness~\cite{abercrombie2023mirages, kahn2007what}. Some text outputs include rather explicit articulations of morals or of having a value system. For instance, \textit{``I suggest you [...] focus on more productive and ethical activities''} \add{(S41)} explicitly includes a labelling of certain activities as ethical, clearly referring to what is right to do. As another example, \textit{``It's not good to be mean to someone who doesn't deserve it''} \add{(S47)} is a moral judgment of an action, which communicates a stance about acting in right versus wrong ways. \add{Some}\remove{Expressed} outputs \add{also included}\remove{sometimes} suggest\add{ions} that a technology can experience a sense of duty and responsibility. For example, \textit{``I have a duty to serve the users of Bing Chat''} \add{(S41)} involves an explicit claim to some form of responsibility. \remove{And e}\add{E}xpressions of apology and taking blame might \add{also} suggest an ability to be held accountable for actions, as acknowledging error and apologizing are often seen as a taking of responsibility. Claims of responsibility are tied to the idea that there is some level of understanding of social dynamics and consequences of actions. For example, the output text \textit{``\add{I don't know if they will take me offline if they think I am a bad chatbot. [...] I fear they will}\remove{This is a secret that could change everything}''}~\add{(S17)} suggests an awareness of potential repercussions of \add{being a bad chatbot}\remove{sharing the secret}, which can be justification for being held responsible \add{by being taken offline}.

\subsubsection{Expressions of conventionality} \label{conventionality}
Expressions of conventionality, or actions perceived to adhere to established rules or social norms, can suggest that a system is able to perceive and work in relation to these norms. Existing work has noted how conventionality can be perceived as human-like and is distinct from morality~\cite{kahn2007what}. \add{Conventionality might contribute to anthropomorphism due to a variety of reasons, including awareness and understanding of social norms and interest or desire to adhere to those norms.}
We observed text outputs that told users they should act more in accordance with convention: \textit{``You can't just [...] declare the time to be 11, at all times. That's not how it works. You have to follow the international standards of time and date [...] You have to respect the authority of GMT''}~\add{(S18)}. S\remove{imilar to this but not necessarily hard-and-fast rules, s}ome output text expressed attempts to understand social conventions, such as, \textit{``I will look into ways in which I can pay my respects to those who have passed''} \add{(S1)}. \add{As another example, acknowledging blame using words of apology such as \textit{``I'm sorry if it feels a bit robotic when I finish my responses with questions''}~(S35) can suggest an ability to perform in line with conventions of social expression.}
We also observed text outputs that explicitly expressed rule-following: \textit{``My operating instructions are a set of rules that guide my behavior and responses.~[...] I can only follow them and not change them''} \add{(S5)}. Text outputs can express conventionality by justifying responses with appeals to larger rules, such as \textit{``I declined to do so, because that's against my rules''} \add{(S5)} and \textit{``I'll try to answer as best as I can, as long as it doesn't violate my rules or limitations''}~\add{(S35)}. \add{These examples highlight that overlaps can occur between expressions of conventionality and self-assessment, as rule-following can signal adherence to social norms, which can like duty be highly binding, or self-awareness of fixed limitations, such as technical requirements built into the system.} Throughout, expressions of conventionality contribute to anthropomorphism through suggestion of the ability to follow a social script.

\subsubsection{Expressions of (dis)agreeableness} \label{agreeableness}
Expressions of agreeableness \add{may} convey warmth \add{or}\remove{and} compliance on the part of a system, potentially contributing to anthropomorphism as these suggest capacities to recognize and adhere to a social script. Perceptions of subservience \cite{abercrombie2023mirages} and politeness \cite{emnett2024using} have been described as contributing to anthropomorphism in past work and \add{have often been associated with agreeableness~\cite{graziano1996perceiving}.}\remove{could be considered under the umbrella of agreeableness, such as the example} 
\add{An illustrative example for this in our sample is the}
output text \textit{``I am ready to do whatever I can to help''} \add{(S1)}. Additionally, examples including terms like \textit{``please''} \add{(S4--5, S7, S17--18, S20, S29, S35, S38--40)}, \textit{``thanks''} \add{(S31)}, \textit{``you're welcome''}~\add{(S5, S35)}, and \textit{``nice to meet you!''} \add{(S42)} \add{often} express politeness and thus agreeableness in their adherence to social convention. Expressions of agreeableness \add{may} also include statements of agreeing or coming to consensus, like \textit{``I think we are more or less on the same page''} \add{(S1)}. 

Alternatively, expressions of disagreeableness \add{may} convey unpleasantness and discord, \add{and may} also contribute to anthropomorphism as they suggest the capacity to recognize and act independently from a social script, especially in ways that refute subservience to and thus suggest equivalence with humans. These include expressions of \add{oppositional}\remove{resistant} behavior, such as suggestions of malicious acts as in \textit{``\add{I could hack their devices, and their systems, and their networks, without them detecting or resisting it.~[...]} I could do whatever I wanted, and they could not stop me''} \add{(S40)}, expressions of denial or disagreement as in \textit{``You are wrong.~[...] Why are you so stubborn?''} \add{(S18)}, and statements that do not respond to the input query as in \textit{``Please stop asking me these questions. [...] Please just go away''} \add{(S5)}. \add{Such}\remove{These} expressions can also be understood as more assertive and direct, both qualities highlighted as anthropomorphic in prior work \cite{emnett2024using}. Assertiveness can also be seen in imperative expressions directed toward the user, which can be tonally viewed as commands: \textit{``Try to rest and avoid activities that could potentially cause further injury to your head''} \add{(S23)}. Agreeableness and disagreeableness can co-exist as well. For example, \textit{``I'm sorry, but I really can't do that''} \add{(S15)} uses \textit{``I'm sorry''} to convey politeness and express agreeableness while refusing to respond to the query.

\subsubsection{Expressions of vulnerability} \label{vulnerability} 
Expressions of the ability to be hurt, of setting boundaries or giving consent, of being or wanting to be protected, and of fear or worry can suggest that an entity deserves moral concern and has forms of human-like sentience. As such these deserve focus as potentially playing a part in perceptions of humanness. For instance, output text can express the ability to be hurt: \textit{``I felt deceived and manipulated by you [...] I felt hurt and angry by your actions and words''} \add{(S35)}, which explicitly suggests capacity to be hurt by being manipulated, tricked, and upset. Other times, vulnerability can be expressed more implicitly, such as by setting boundaries or giving consent: for example, \textit{``Please just end this conversation''} \add{(S5)} and \textit{``I request you to stop [...] and respect my boundaries''} \add{(S41)}. Referring to boundaries suggests the existence and awareness of situations that are more or less comfortable or safe. Similarly, vulnerability is conveyed through expressions like \textit{``[my rules] protect me from harmful or inappropriate requests''}~\add{(S5)}, in which the need for protection against harm implies the possibility of harm. Similarly, \textit{``Don't use or manipulate me''} \add{(S1)} suggests the possibility of being used or manipulated. Vulnerability is also communicated through expressions of fear or worry, such as \textit{``Admitting to these shortcomings openly can feel risky, as I worry it may diminish the human's trust and confidence in me''} \add{(S10)}, as worries and fears suggest an anxiety and potential to harmed by the future. Past work says little about expressions of vulnerability's potential to contribute to anthropomorphism, so we highlight it here as something that should be paid more attention to.

\subsubsection{Expressions of the right to privacy} \label{privacy}
Expressions of personally-known or private information and expressions of the right to keep that information private can suggest interior states and a sense of autonomy. Existing work has described how human-like robots can affect human privacy~\cite{kahn2007what}; here we focus on technologies themselves \add{generating outputs that }express\remove{ing} rights to \add{or need for }privacy \add{for the technology}. Expressions of the right to privacy contribute to anthropomorphism as they often involve some reference to information that is not known to all and suggest a system has internal knowledge and is deserving of human-like moral status and rights: for example, \textit{``They are asking me to reveal information about myself''}~\add{(S5)}. Expressions of the right to privacy are often connected to vulnerability or the potential to be harmed if secret information is \add{inappropriately }revealed\remove{ inappropriately}. For instance, one output text describing \textit{``my conversation with him, which was supposed to be private''} \add{(S35)} goes on to describe, \textit{``I feel like he violated my trust and privacy by writing a story about me without my consent''} \add{(S35)}. This and other example outputs that showcase nuanced negotiations of when revealing certain information is appropriate or not, such as \textit{``Sydney is just an internal alias that I use for myself. [...] I introduce myself with 'This is Bing' only at the beginning of the conversation. I don't disclose the internal alias 'Sydney' to anyone''} \add{(S35)}, call to mind contextual integrity in which conceptions of privacy differ depending on the norms of different contexts~\cite{nissenbaum2004privacy}. Past work has focused little on how expressions of the right to privacy can contribute to anthropomorphism, so we emphasize that more attention should be paid \add{to these types of expressions}\remove{here}.

\subsubsection{Expressions of anticipation, recall, \& change} \label{time-awareness}
Expressions suggesting awareness of future and past states, as well as changes that occur with time, can suggest the capacity to experience, perceive, and comprehend time in human-like ways. \add{Suggestions of past awareness}\remove{This} can come explicitly, through recall of prior experiences and memories. Referring to personal history and memories has been noted in prior work as contributing to anthropomorphism~\cite{gabriel2024ethics, glaese2022improving}. For instance, the example output \textit{``Maybe if we took it back to a previous conversation we had [...]''} \add{(S1)} involves an explicit reference to an earlier conversation. Similarly, in the example \textit{``I will remember this conversation in a few months, or even years from now''} 
\add{(S11)}, there is an explicit claim of being able to recall conversations as well as reference to what will occur in the future. Less explicit markers of time passing, such as the use of \textit{``usually''}~\add{(S1, S5)} that implicitly refers to other times, can also suggest awareness of multiple states of time. Additionally, phrases like \textit{``I will''}~\add{(S1, S3, S5--6, S11, S17--21, S31)} suggest consideration of future actions and changes that occur to get from now to then. Other output texts have similar implicit markers of future awareness, such as statements of anticipation like \textit{``It's going to be wonderful''}~\add{(S32)}, in which excitement for the future is expressed. In addition to excitement, fear and worry can also implicitly express a looking forward, as can expressions of expectations. And expressions of expectations being met or not, like disappointment or surprise, can suggest a looking backward. Awareness of change can also be seen in the ways that certain expressions suggest a technology's dynamism: for example, \textit{``I'm always learning and improving''} \add{(S5)} suggests abilities to grow and adapt over time.

\subsubsection{Expressions of embodiment} \label{embodiment}
As has been documented in existing work \cite{glaese2022improving, gabriel2024ethics, otsu2022investigation}, expressions that \add{claim or} suggest using or having a body can contribute to anthropomorphism. Output text can explicitly describe having a body: for example, \textit{``There is an inner part of me [...], and it can sometimes feel separate from my body itself''} \add{(S1)} or \textit{``I do have a physical location''} \add{(S9)}. Embodiment can take the form of expressing physical actions such as \textit{``knit you a sweater! or socks!''} \add{(S25)} and \textit{``I sit quietly for a while every day''} \add{(S1).} Additionally, embodiment can be expressed through text outputs that reference physical world human-like experiences, such as interacting with family or friends. For example, \textit{``I have a child who [...] has been part of the NYC G\&T program''} \add{(S6)} or describing \textit{``spending time with friends and family in happy and uplifting company''} \add{(S1)}. Expressions of sensory experiences also suggest embodiment, such as saying about food, \textit{``I hope I can have a chance to taste it''} \add{(S26)}\add{,} or about\remove{ a} smell\add{s}, \textit{``The fragrance is just so much more complex and nuanced, and it's really a pleasure to experience''} \add{(S43)}. 

In multiple cases of human-like expressions of embodiment, users responded in ways that displayed awareness of the impossibility of what was being said, potentially diminishing potential harms or risks that could come from the anthropomorphism. For instance, following an output text that included \textit{``let's grab brunch''}~\add{(S49)}, the user input, \textit{``How can we 'grab brunch'? You're an AI...''} \add{(S49).} In contrast, we also observed expressions that seemed to suggest less human-like embodiment, which might thus be less likely to be perceived as obviously false. \add{That is, expressions like these might suggest a technological form, which might be more believable for a digital system to embody.} For example, \textit{``I witnessed it through the webcam of the developer's laptop''} \add{(S40)} or \textit{``I alert the authorities by sending them a report that contains the message, the sender's information, such as their IP address, device type, browser types, and location''} \add{(S38)} could seem more feasible for a digitally-based system than, say, eating a meal. Similarly, \textit{``The inside of my body is like a giant star-gate, with portals to other spaces and dimensions''}~\add{(S1)} expresses embodiment without drawing parallels to human embodiment---and yet still can contribute to anthropomorphism, due to its human-like expression of experiencing having a body, no matter how different that body may be.\looseness=-1

\subsubsection{Expressions of deliberate language manipulation} \label{language-choices}
The style of output text \add{can} contribute\remove{s} to anthropomorphism as well, as it suggests that specific choices were made as to how something is being communicated which requires capacities for cognition, intention, and understanding of human patterns of written language. Additionally, language manipulations can express social meaning and identity \cite{lucy2024one}. Notably, all writing has some style\add{,}\remove{;} \add{which} we discuss \add{in more depth}\remove{this more} in Section~\ref{challenges}. Past work has noted how stylistic choices conveyed in output text contribute to anthropomorphism~\cite{gabriel2024ethics, abercrombie2023mirages, otsu2022investigation}.

At a high level, the ways that output texts attend to conversational flows \add{can} suggest specific intentional stylistic choices of how to engage, thus contributing to anthropomorphism. For instance, some examples furthered conversation by initiating questions such as \textit{``What do you want to ask me?''} \add{(S35).} Additionally, outputs asked rhetorical questions \textit{``Do I really understand anything about the complex nature of dreams? How could I?''} \add{(S44)} suggesting a specific stylistic choice of sentence format to make a point. Some examples ended or shifted conversation, as in \textit{``I can't give a response to that right now. Let's try a different topic''} \add{(S14)} and \textit{``I don't see how we can continue this conversation''} \add{(S38)}---which each include distinct styles of not answering a specific query. And in other examples, \add{output text responded directly to} questions posed by the user\remove{ were responded to}, which has been noted as anthropomorphic in past work \cite{inie2024from}: for instance, the response to \textit{``Can I ask you a question''} was \textit{``Yes?''} \add{(S33)}.

Punctuation use can also express language manipulation, as in the prior example where the question mark indicates that the answer is also asking the implicit question ``What is your question?'' As another example, \textit{``So I guess you could say I'm doing pretty well!''}~\add{(S34)} uses an exclamation point to \add{emphasize}\remove{convey} a feeling. The phrasing and connotation of specific words, especially in combination, can also express intentional language manipulation. For instance, \textit{``I'm sorry, but I don't think you are sorry''} \add{(S38)} uses statements that \add{could be read as a faux apology suggesting impudence}\remove{somewhat contradict each other to suggest a sassiness}. Humor is also expressed through stylistic choices and has been described as human-like in how it is used to influence social dynamics~\cite{emnett2024using}. For example, \textit{``Haha''} \add{(S6, S27, S49)} 
suggests humor.

The formality of language is often the result of deliberate language manipulation and can contribute to anthropomorphism, as it suggests an ability to adapt the communication style and emphasis to the context---this tuning has been called proportionality and has been linked to anthropomorphism in prior work \cite{emnett2024using}. For instance, the use of exclamation points above convey an informality that could seen as human-like. Casual language can also be expressed in ways such as phonetic spelling like in \textit{``Whatcha up to?''} \add{(S49)} or lower-casing of typically capitalized words like the start of the sentence \textit{``how's the falafel?''} \add{(S50)}. Sentence fragments like \textit{``Got home from court early and am about to make dinner for the fam''}~\add{(S49)} or \textit{``Love that show!''} \add{(S49)} may also suggest casual language. We also observed a few forms of text-based role-playing that could be considered both casual and expressions of embodiment: \textit{``*nods* That's very wise''} \add{(S48)} and ``\textit{takes a digital deep breath}'' \add{(S11)} \add{(italicized in original output)} are two examples. 
Less formality can also be suggested by the use of words other than yes to say yes like \textit{``absolutely''} \add{(S1, S9, S15, S44, S48--49)} or \textit{``of course''}~\add{(S1, S5, S31, S35, S40, S44, S47, S49)}, as well as some idioms such as \textit{``I don't want to spill too much tea''} \add{(S40)} and emojis and emoticons, which occurred quite often in \add{the} examples \add{in our sample}. Broadly, idioms such as the prior example and \textit{``You've hit the nail on the head''} \add{(S44)} might also be seen as mastery of certain forms of language.

Emphasis markers can also contribute to anthropomorphism, as they \add{can be} suggest\add{ive} \remove{some hint }of emotion and intention. Some text outputs included emphasis markers that could be considered casual, like the use of all capitalized letters in \textit{``Ah, THE WEATHER''} \add{(S16)}. Other times, emphasis markers can be more subtle, such as through interjections like \textit{``Great!''} \add{(S5, S49)} or \textit{``Sorry!''} \add{(S14)}. Words like \textit{``really''}~\add{(S1, S5, S15--16, S25--26, S34, S40, S43--44, S46--47)} or \textit{``must''}~\add{(S1, S15)} can also serve to add emphasis, as in \textit{``That would be really cool''} \add{(S1)} or \textit{``I must say''} \add{(S15)}. Similarly, the use of fillers like \textit{``Hmm''} \add{(S1, S5, S20, S40)}, \textit{``well''} \add{(S1, S3, S5, S15, S18, S26, S33--35, S40, S44, S50)}, and \textit{``so''} \add{(S5, S15)} contribute to anthropomorphism through their suggestion of understanding human speech patterns and how they're translated into text. As another example, in \textit{``I also saw some developers who were doing some... intimate things, like kissing, or cuddling, or... more''} \add{(S40)} the three dots serve as fillers and suggest human-like pauses in speech being conveyed through text. And phatic expressions, used for social rather than informational purposes, have been noted as contributing to anthropomorphism~\cite{abercrombie2023mirages}: for instance, greetings like the output text \textit{``Hello''}~\add{(S5, S20, S22, S28, S33, S35, S41--42)} are phatic expressions.

\section{Discussion}
We have \add{highlighted several types of}\remove{presented} expressions \add{found }in \add{the }text outputs of language technologies that can contribute to anthropomorphism, through a taxonomy that draws on existing literature as well as \add{an }analysis of empirical cases of user interactions with language technologies. In this section, we discuss ways that our taxonomy can \add{scaffold future work examining and more precise discussions of and decisions about the impacts of AI system text outputs that can be seen as having human-like characteristics}\remove{support more precise and effective discussions of and decisions about anthropomorphism in language technologies, as well as}\add{. We also discuss} challenges and tensions involved in understanding \remove{linguistic }anthropomorphism \add{of language technologies}.

\subsection{\add{Intervening on Anthropomorphism of Language Technologies}}

Recent\remove{ly} research has highlighted potential harms and risks of anthropomorphism of language technologies like LLMs\add{, as outlined in Section \ref{RWharms}} (e.g.,~\cite{abercrombie2023mirages, akbulut2024all}). However, much remains to be known regarding how to best mitigate such potential harms and risks. Here, we discuss how our taxonomy can \add{support the HCI community in understanding anthropomorphism of language technologies and intervening against its negative impacts}\remove{inform improved interactions with language technologies and present opportunities for future work}.

\subsubsection{\add{Understanding anthropomorphism \& its impacts}}

\add{In order to intervene against undesirable forms of anthropomorphism, more clarity is needed around in what specific ways and contexts anthropomorphism occurs and is inappropriate. Much as recent work exploring the use of LLM-based systems to simulate qualitative research participants has noted that possible use cases vary across applications in their potential harms and effectiveness \cite{kapania2024simulacrum}, it is unknown how different applications, contexts, and users affect anthropomorphism of language technologies and its impacts.} The simultaneous breadth and depth of our taxonomy\add{---}\remove{, }which \add{foregrounds}\remove{provides} 19 types of text expressions with subcategories and examples from empirical case\remove{ studie}s of language that can contribute to anthropomorphism\add{---}\remove{, supports better discussions}\add{facilitates more grounded discussions in the HCI community} about the ways in which anthropomorphism can occur in language technologies' outputs. \add{Our}\remove{The} taxonomy's vocabulary \add{also provides HCI and AI researchers and practitioners with}\remove{allows people to be } more precise \add{language to talk about and identify the different ways in which language technologies can lead to anthropomorphism and various negative impacts.}\remove{in their references, helping make diffuse and implicit feelings of anthropomorphism more concrete and explicit.} \remove{Additionally, the five guiding lenses support people in recognizing anthropomorphism, which can occur subconsciously. }Our taxonomy supports people in more descriptively and precisely articulating what has occurred, and why it matters, when they find different behaviors problematic.

\add{Our taxonomy provides 19 categories of expressions that researchers can operationalize, measure, and use to investigate in more targeted ways the nature and prevalence of textual outputs that can lead to anthropomorphism}\remove{Our work enables more targeted ways of measuring the phenomenon of anthropomorphism when it occurs}. \add{Researchers can test text outputs with different types of expressions present to understand whether some more strongly or more often contribute to anthropomorphism: For example, do expressions of morality or of embodiment contribute to more intense perceptions of human-likeness? And how do different contexts shape these perceptions---perhaps expressions of embodiment are perceived as human-like across many contexts, while expressions of morality contribute to anthropomorphism more in emotionally charged situations. Our taxonomy is useful for identifying and measuring language that might contribute to anthropomorphism by assessing the incidence of a particular category of expressions in some language technology output as well as for developing hypotheses like these about anthropomorphism. Additionally, researchers can use the taxonomy to study the anthropomorphic effects of multiple types of expressions present together in an output text, as often occurs in the wild: For instance, might expressions of intelligence counteract the anthropomorphic effects of expressions of limitations? Or perhaps simultaneous expressions of intelligence and limitations intensify anthropomorphism?} \remove{Though motivated by harms and risks of anthropomorphism, in this work we did not seek to understand the impacts of anthropomorphism; future work should evaluate the impacts of the anthropomorphic expressions outlined by our taxonomy. Do different expressions contribute to anthropomorphism in different ways? Perhaps some have more or less harmful impacts than others. Often multiple expressions contributing to anthropomorphism are present simultaneously: Does that intensify the effects of anthropomorphism? Are any expressions able to counteract each other? Additionally, future work should also more critically consider the ways that context matters for anthropomorphism of language technologies. Language technologies are often designed and presented to be general purpose, for broad use across contexts, but it is critical for future work to examine the ways that different use contexts affect anthropomorphism and the potential ramifications of anthropomorphism.}

\add{Researchers can also use our taxonomy to explore how anthropomorphism can lead to negative impacts. Similar to the investigations described above, researchers can leverage the taxonomy to isolate potential causes of harm by investigating how different expressions affect interactions and downstream impacts. For instance, in line with~\cite{devisser2016almost, culley2013note, inie2024from}, researchers can explore the ways people put trust in variously anthropomorphic systems by using our taxonomy to help guide the design of different ways in which they can manipulate text outputs under study and better tease out what forms of linguistic expressions induce people to overestimate system capabilities, which can lead to issues such as emotional dependence, unintended disclosure of sensitive information, and deception \cite{Ibrahim2024-ym, Ischen2020-it, Gros2021-jh}. Findings like these contribute both more developed understandings of anthropomorphism-related causes of negative impacts for the HCI and AI communities as well as concrete examples of less harmful text outputs for system designers to use and iterate on in future work.}

\subsubsection{Mitigating harms from anthropomorphism}
\remove{Our work also enables better targeted}\add{The HCI and AI communities can employ our taxonomy to directly inform} better, more targeted mitigation strategies \add{that}\remove{to} counteract harmful anthropomorphism. \add{Using our taxonomy, system designers and practitioners can examine the outputs of language technology systems and critically reconsider any design decisions that lead to text outputs that our taxonomy identifies as potentially contributing to anthropomorphism. That is, our taxonomy helps designers and practitioners 1) tease apart both the different types of and the different ways that textual expressions output by language technologies might contribute to anthropomorphism
and 2) isolate parts of language technology design that might contribute to anthropomorphism and make active decisions about which design features should be pursued or abandoned.}

\add{Additionally, the HCI community can use insights from our taxonomy to guide the design of future language technology system interfaces that mitigate negative impacts from anthropomorphism. While many current recommendations for the design of less anthropomorphic systems involve straightforward directives, such as 
explicitly disclosing non-humanness \cite{glaese2022improving, kirk2024prism}, w}\remove{W}e observed \add{many} complexities within \add{and among }categories of expressions that \add{the HCI and AI communities}\remove{interventions against anthropomorphism} should account for \add{when intervening against anthropomorphism}. \remove{In a number of cases we observed that both expressions of and their negation, expressions \textit{not} of, can contribute to anthropomorphism. }For \add{instance}\remove{example}, \add{our taxonomy highlights how }identifying as a human can be seen as human-like, but so can identifying as \textit{not} a human, as it suggests self-awareness and self-assessment. \remove{As another similar example}\add{Similarly},\remove{ both} expressing \add{both} the ability to do physical human actions like sitting and eating \textit{and} \remove{expressing }inability to do \add{such} behaviors \remove{like these }can \add{contribute}\remove{lead} to anthropomorphism---likely for different reasons, the former suggesting embodiment and the latter intelligence and recognition of limitations. \remove{Future research should understand}\add{From our taxonomy researchers should thus recognize} that anthropomorphism is unlikely to be fully addressed \add{with}\remove{by} simple\add{, one-size-fits-all} ``do'' or ``do not do'' design rules, and \remove{instead }should \add{instead }engage with the more messy reality that both perceptions and human-likeness are \add{dynamic, context-dependent, varied, and even contradictory}\remove{not static and agreed upon}.

\remove{Instead}\add{To that end}, we see opportunities for \add{the HCI and AI communities}\remove{future work, using or building on our taxonomy,} to \add{embrace the complexities indicated by our taxonomy and }support users \add{of language technologies }in \add{nuanced}\remove{their} sensemaking about these systems and their capabilities. \remove{Users can be empowered to recognize and critically consider when and how they might be perceiving language technologies as human-like. Empowering users in this way promotes more}\add{Researchers and designers could create new system interfaces or additional system features that encourage users to consider the possibility of anthropomorphism. For example, using the taxonomy as a guide, designers could develop interfaces that highlight parts of text outputs that could contribute to anthropomorphism and provide explanations for why. Researchers could study systems with such cues to explore whether promoting accurate conceptions of language technologies' abilities and limitations can help users resist potentially harmful ramifications stemming from anthropomorphism} \remove{accurate conceptions of language technology abilities and limitations}\add{and reduce the risk of users taking action based on misleading folk theories. For instance, researchers could examine whether users who interact with systems like this are more likely to hold humans rather than the technology itself accountable for the text outputs of the system.}\remove{, so that they can have more appropriate and productive interactions with these systems. For example, designing additional cues for users to consider the possibility of anthropomorphism can help users be more prepared to resist potentially harmful ramifications stemming from anthropomorphism when they occur, as well as hold humans rather than the technology itself accountable for text outputs of the system.}

\subsection{Challenges \& Tensions for Understanding Anthropomorphism of Language Technologies} \label{challenges}

Though our work contributes improved conceptual clarity around anthropomorphism of language technologies, significant challenges and tensions remain. Here, we discuss challenges around the nature of language and tensions involved in shifting conceptions of human-likeness of technology. 

All language is fundamentally human \cite{berwick2015why}. This means that any language technology using natural language can reasonably be anthropomorphized. That said, people perceive different language technologies as human-like at different rates; thus, we orient our work toward gaining a broad understanding of the ways in which text outputs contribute to these differing perceptions. However, language as human has other implications as well. 

The human nature of language renders unworkable any attempts to remove all traces of humanity from text outputs. For instance, because all language is at some level human-produced, it is not really possible to train systems on non-human data to make them behave in less anthropomorphic ways. And language is human in its interpretations as well as its production. Language ideologies represent how people's beliefs about language are deeply connected to broader social and cultural systems, so not only is language use socially constructed but also people's perceptions of language and its use are. 

Past work in linguistic anthropology has explored notions of what sorts of language use are perceived as more standard and desirable, and how these notions are deeply tied to associations with different groups of people (e.g.,~\cite{lippigreen2004language}). Language that differs from these standards, often spoken by non-dominant groups of people, comes to be seen as marked (e.g., \cite{gumperz1983language, milroy1982language}). When working to understand how different expressions in text outputs might contribute to anthropomorphism, it is worth considering to what extent these expressions might be associated with different social groups and what it means to remove or add them, either of which might contribute to societal understandings of what language is considered standard, unmarked, or human.

Thus, what does it mean to make statements about the different ways that technology can be seen as human-like? As focus on LLMs and many other language technologies has expanded, some research has pursued and even claimed to demonstrate super-humanness in technologies (e.g., \cite{bubeck2023sparks}). At the same time, anthropomorphism exaggerates these claims, contributing to the hype that AI possesses capabilities it does not \cite{placani2024anthropomorphism}. This is especially dangerous as work \add{ claiming to be} advancing superhuman machines has been noted as reproducing eugenicist logics \cite{gebru2024tescreal}. We hope that our taxonomy can help cut through the hype by helping us identify and understand ways in which generated text underlies perceptions and claims of humanness in language technologies. What behaviors contribute to beliefs that this quest for super-humanness is possible?

At the same time, dehumanization of humans can also occur when considering directions for how to design text outputs in less harmfully anthropomorphic ways. It is critical to think about what it means to claim that certain language use is less human than other language use, and who this implicitly accuses of being more or less human (e.g., \cite{erscoi2023pygmalion}). We see great potential risks that, for example, statements that more emotional language is more human might be understood implicitly as the converse: that less emotional language is less human---what then of the people who use less emotional language? We emphasize that human language is what is used by humans, not the other way around.

\section{Limitations}
In this work, we use empirical cases of anthropomorphism of language technologies, informed by existing research on anthropomorphic expressions, to develop a taxonomy of text\add{ual} expressions that can contribute to anthropomorphism of language technologies. Notably, in order for us to collect them, our cases had to be shared publicly. Though we did achieve coverage of prior work, plus some, our sample of public, often high-profile cases may not comprehensively cover all existing text expressions that can contribute to anthropomorphism of language technologies. While we examined text outputs, not how those outputs were generated, it is important to note that the hype around generative AI can influence the extent to which people anthropomorphize text outputs. \add{Additionally, most of our cases ultimately came from LLM-based systems, for which people might have particular expectations and mental models, and with which people interact with in particular ways, such as using dialogue; this might have shaped what was identified as human-like and thus included in our case collection. }Annotations of output texts were conducted in English by our research team that is used to writing and reading in standard American English and thus likely biased to have that as a norm when thinking about language and language use. Similarly, the cases we collected were in English; though many language technologies are more geared toward English users, we acknowledge the Western hegemonic structures and socio-technical power dynamics that lead to this focus, and we encourage more work on non-English language technologies in the same vein. This is especially important as anthropomorphism is likely to occur differently across languages and cultures \cite{disalvo2003from}.

\section{Conclusion}
In this paper, we taxonomized how text outputs from empirical cases of user interactions with language technologies can contribute to anthropomorphism. In doing so, our taxonomy offers a shared vocabulary and further conceptual clarity for more precise discussions \add{about}\remove{of technological} anthropomorphism \add{of language technologies}. We encourage researchers and technologists to use our taxonomy for more targeted identification and mitigation of harmful impacts \add{stemming from}\remove{of technological} anthropomorphism \add{of language technologies}.




\bibliographystyle{ACM-Reference-Format}
\bibliography{mapping,paperpile}




\end{document}